\documentstyle[floats,aps,epsf,prb]{revtex} % PREPRINT format
\begin{document}
\draft

\twocolumn[\hsize\textwidth\columnwidth\hsize\csname
@twocolumnfalse\endcsname

%...................................................................
%
%
\title{Core reconstruction of the 90$^\circ$ partial dislocation in
non-polar semiconductors}

\author{R.W. Nunes$^1$, J. Bennetto$^2$, and David Vanderbilt$^2$}
\address{$^1$Complex System Theory Branch, Naval Research Laboratory,
Washington DC, 20375-53459\\
and Computational Sciences and Informatics, George Mason University,
Fairfax, Virginia\\
$^2$Department of Physics and Astronomy, Rutgers University,
Piscataway, New Jersey 08854-8019\\}

\date{\today}
\maketitle
%.....................................................................

\begin{abstract} 
We investigate the energetics of the single-period and double-period
core reconstructions of the 90$^\circ$ partial dislocation in the
homopolar semiconductors C, Si, and Ge. The double-period geometry is
found to be lower in energy in all three materials, and the energy
difference between the two geometries is shown to follow the same
trends as the energy gap and the stiffness. Both structures are fully
reconstructed, consisting entirely of fourfold coordinated
atoms. They differ primarily in the detail of the local strains
introduced by the the two reconstructions in the core region. The
double-period structure is shown to introduce smaller average
bond-length deviations, at the expense of slightly larger average
bond-angle bending distortions, with respect to the single-period
core.  The balance between these two strain components leads to the
lower energy of the double-period reconstruction.
\end{abstract}

%.....................................................................
\vskip2pc]

\narrowtext

A fundamental understanding of plasticity in solids clearly requires a
knowledge of the atomistic structure at the cores of dislocations.  In
particular, the microscopic mechanisms of dislocation motion are
intimately related to the defects (e.g., kinks) that can occur within
the dislocation, which in turn are connected with the underlying
lattice symmetries and the nature of the reconstruction in the
core.\cite{bulatov1} Recently, advances in computer power and
computational methodology have led to an active area of research
focused on the theoretical study of the atomistic structure of these
dislocation cores and their defects.\cite{nbv1,bnv,nbv2,bulatov2,%
bigger,hansen,arias,jones79,markl79,markl83,chelik,lodge,markl92,%
markl94,jones80,jones93,heggie83,heggie87,heggie93,oberg}
However, in at least one important case, even the structure of the
dislocation core itself remains fundamentally in doubt.

In tetrahedrally bonded semiconductors, the two most frequently
occurring dislocations are the 30$^\circ$ and the 90$^\circ$ partial
dislocations, lying on \{111\} planes along [110]
directions.\cite{alexan,duesbery,hirsch} These materials are of
obvious technological importance, and the detailed understanding of
the atomic structure at the dislocation cores is of great interest,
since dislocations influence both the electronic and the mechanical
properties of semiconductor devices. Indeed, a great deal of
theoretical effort has been devoted to study the properties of
dislocations in these
materials.\cite{nbv1,bnv,nbv2,bulatov2,bigger,hansen,arias,jones79,%
markl79,markl83,chelik,lodge,markl92,markl94,jones80,jones93,heggie83,%
heggie87,heggie93,oberg} In the particular case of silicon, the
theoretical study of the dislocation cores at the atomistic scale has
revealed a rich structure of point excitations (kinks and
reconstruction defects) in the core of the 30$^\circ$ and 90$^\circ$
partials.\cite{nbv1,bnv,nbv2,bulatov2,jones80,jones93,heggie83,heggie87,%
heggie93}

\begin{figure}[!t]
\epsfxsize=2.0in
\centerline{\epsfbox{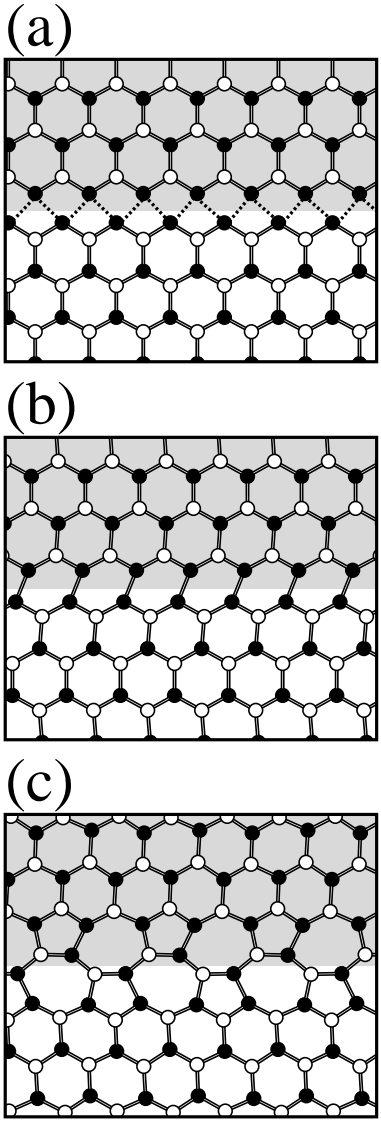}}
\bigskip
\caption{(a) Symmetric reconstruction of the 90$^\circ$ partial
dislocation in homopolar semiconductors. Shaded area indicates
stacking fault. (b) The single-period (SP) symmetry-breaking
reconstruction.  (c) The double-period (DP) symmetry-breaking
reconstruction.}
\label{fig1}
\end{figure}

Most of these works have concentrated on understanding the structure
of the 90$^\circ$ partial dislocation. Until recently, a consensus had
emerged at the theoretical level, about the nature of the
reconstruction at the core of the defect. In its unreconstructed
configuration, the core of the 90$^\circ$ partial displays a zigzag
chain of threefold-coordinated atoms running along the dislocation
direction, with broken bonds lying nearly parallel to the slip plane.
Mirror symmetry planes are present is this
configuration, as can be seen in Fig.~\ref{fig1}(a).  A variant of
this structure is one in which the dashed lines in Fig.~\ref{fig1}(a)
are considered to be covalent bonds, resulting in a ``quasifivefold''
reconstruction that also retains the mirror symmetry.\cite{bigger} On
the other hand, a reconstruction that breaks the mirror symmetry of
the unreconstructed core, while preserving the lattice periodicity
along the line, is shown in Fig.~\ref{fig1}(b). In this case all
dangling bonds have been eliminated, and all the atoms are fourfold
coordinated.  Such a reconstruction was predicted to be substantially
lower in energy than both the unreconstructed and the
quasifivefold-reconstructed cores,
\cite{nbv1,bigger,markl83,chelik,lodge,markl94} and thus to be the one
expected to occur in nature.  We will refer to this symmetry-breaking
reconstruction as the single-period (SP) reconstruction.

However, our recent theoretical work on the core
reconstruction of this dislocation in Si produced a surprise.  In
Ref.~\onlinecite{bnv}, we proposed an alternative core structure for
the 90$^\circ$ partial in which, in addition to symmetry breaking of
the SP core, the periodicity along the dislocation line is
doubled. This double-period (DP) reconstruction, which is shown in
Fig.~\ref{fig1}(c), can be derived from the SP one by introducing
alternating kinks at every lattice site along the core. This geometry
is consistent with all available experimental information about the
90$^\circ$ partial. Like the SP core, the DP structure is fully
reconstructed, and thus neither one gives rise to deep-gap states
which would show an EPR signal. EPR experiments in Si indicate a very
small density of dangling bonds in the core of
dislocation.\cite{alexan,duesbery,hirsch} Moreover, both cores consist
entirely of fivefold, sixfold, and sevenfold rings, both being
consistent with images produced by transmission electron microscopy,
at the current level of resolution of this technique.\cite{kolar}
The results we
obtained in Ref.~\onlinecite{bnv} show the DP structure to be lower in
energy than the SP one, by means of Keating-potential, total-energy
tight-binding (TETB), and {\it ab initio} local-density (LDA)
calculations.

In the present work, we investigate the issue of DP versus SP
reconstruction in the homopolar diamond-structure semiconductors
C, Si, and Ge.  As in our previous work on Si, we present
LDA, TETB, and Keating-potential results for the energies of the two
core reconstructions. The trend in the energy difference between the
DP and SP cores is seen to correlate with the stiffness of each
material, with the DP core being more (less) strongly favored in C
(Ge), as revealed by our {\it ab-initio} and TETB calculations. This
is in contrast with the Keating-potential calculations which, despite
predicting the correct ground-state and trend between Si and Ge, favor
the SP structure over the DP core in the case of C.

As in Ref.~\onlinecite{bnv}, we employed supercells of 96 and 192
atoms for the SP and DP cores respectively, containing a dislocation
dipole in the quadrupole arrangement suggested in
Ref.~\onlinecite{bigger}. These were fully relaxed (with average
forces no larger than 0.01 eV/\AA\ in each case) using an LDA
approach, with core states represented by Troullier-Martins
pseudopotentials, as implemented in the fhi96md package.\cite{fhi}
All the energy differences were converged to within $\sim$5\% with
respect to plane-wave cutoff. For Si and C, these cells were also
relaxed with a TETB model, using an ${\cal O}$(N) density-matrix
technique\cite{lnv} to solve for the electronic structure.  (The
corresponding TETB results for Ge would be expected to resemble those
for Si.)  The tight-binding Hamiltonian of Kwon {\it et al.}\cite{kwon}
was used
for Si, while the model proposed by Xu {\it et al.} was applied to
C.\cite{xu} The convergence of the TETB results with respect to cell
size was checked by relaxing larger cells, with 576 (288) atoms for
the DP (SP) structure.  Keating-potential results were used in order
to investigate qualitatively the strains associated with the two
structures. For these, we use the original set of parameters
introduced by Keating\cite{keating} for the three materials.
Further details about the supercells and the technical aspects of our
TB calculations can be found in Refs.~\onlinecite{nbv1,bnv,nbv2}.

\begin{table}
\caption{Calculated energy difference in meV/\AA, between the SP-
and DP-core reconstructions of the 90$^\circ$~partial in C, Si, and
Ge. Cell size refers to the double-period cell.  $E_{\rm DP}$ is the
energy of the double-period reconstruction.  For the single-period
case, $\overline{E}_{\rm SP}$ and $\Delta E_{\rm SP}$ are respectively
the average and difference of the energies for the two different
relative arrangements of mirror symmetry-breaking.}
\smallskip
\begin{tabular}{ldddd}
 &\multicolumn{2}{c}{192-atom supercell}
 &\multicolumn{2}{c}{588-atom supercell}\\
 &$E_{\rm DP} - \overline{E}_{\rm SP}$ &$\Delta E_{\rm SP}$
 &$E_{\rm DP} - \overline{E}_{\rm SP}$ &$\Delta E_{\rm SP}$\\
\hline
C \\
\quad LDA             &-235   &126    \\
\quad TETB            &-100   & 74     &-76    &14     \\
\quad Keating         & -21   &123     & 34    &24     \\
\quad Keating\tablenotemark[1]
                      &-121   &160    \\
Si \\
\quad LDA             & -69    &48     \\
\quad TETB            & -75    &39     &-57    & 3     \\
\quad Keating         & -27    &40     & -7    & 8     \\
\quad Keating\tablenotemark[1]
                      & -40    &67     \\
Ge \\
\quad LDA             & -58    &27     \\
\quad Keating         & -21    &32     & -5    & 6     \\
\quad Keating\tablenotemark[1]
                      & -12    &36     \\
\end{tabular}
\tablenotetext[1]{Evaluated at LDA-relaxed structure.}
\end{table}

We discuss first the LDA results for the 192-atom supercell as shown
in Table I.  The calculated energy of the SP core depends on whether
the breaking of mirror symmetry occurs with the same sense, or with
opposite sense, for the two dislocations in the supercell.  (There is
a corresponding energy splitting in the DP core, but this is of a much
smaller magnitude and was not taken into consideration.)  As in
Ref.~\onlinecite{bnv}, the average energy of these two possibilities is
denoted by $\overline{E}_{\rm SP}$, while the difference is denoted by
$\Delta E_{\rm SP}$; the latter is expected to vanish in the limit
that the supercell gets large.  Focusing for the moment on the
$\overline{E}_{\rm SP}$ values, it can be seen that the DP structure
is preferred by a substantial margin for all three, and that this
preference follows the same trends as the stiffness and the size of
the band gap of the material.  That is, C shows the strongest tendency
toward stabilization of the DP core, followed by Si and then Ge.

Since we cannot easily afford to repeat the LDA calculation for a
larger supercell, we have carried out parallel calculations on
192-atom and 588-atom supercells of C and Si using the TETB
method. Table I shows first of all that the TETB results are in
qualitative agreement with the LDA ones for both materials,
although in the case of C the corresponding values are
underestimated by a factor of about two.  Secondly, the TETB
results give us a good estimate of the importance of the finite
supercell-size effects.  That is, by inspecting the TETB results
for the two supercells, it appears that one can say that
$\overline{E}_{\rm SP}$ gives the energy of the true isolated SP
structure to within an error bar of approximately $\pm\Delta E_{\rm
SP}/2$.  Applying this heuristic to the LDA results, we see that
the supercell-size error is almost certainly insufficient to
reverse the sign of the predicted $E_{\rm DP}-E_{\rm SP}$.
Returning now to the discrepancy between the TETB and LDA results
for C, this could be attributed at first sight
to the fact that the model of Xu {\it et al.} that we use for C
results in elastic constants which are too
soft [$(C_{11}-C_{12})$ and $C_{44}$ are smaller by 35\% and 17\% with
respect to experimental values, respectively]. On the other hand, the
TETB results for Si that we obtained using the model of
Ref.~\onlinecite{wang}, which also underestimates the elastic
constants, are practically identical to those shown in Table I for
the Kwon model.

\begin{table}
\caption{Minimum, maximum, and root-mean-square variations of bond
lengths and bond angles for the LDA-relaxed SP and DP structures,
relative to the corresponding bulk diamond values.}
\smallskip
\begin{tabular}{lrrrr}
 &\multicolumn{2}{c}{bond length}
 &\multicolumn{2}{c}{bond angle}\\
 & \multicolumn{1}{c}{SP}&  \multicolumn{1}{c}{DP}
 & \multicolumn{1}{c}{SP}&  \multicolumn{1}{c}{DP} \\
\hline
C \\
\quad min          &$-$5.3\%   &$-$4.4\%   &$-$11\%    &$-$14\%     \\
\quad max          &+5.4\%   &+6.2\%   &+20\%    &+22\%     \\
\quad rms          & 3.1\%   & 2.8\%   &3.4\%    &3.6\%    \\
Si \\
\quad min          &$-$2.2\%   &$-$2.1\%   &$-$11\%    &$-$15\% \\
\quad max          &+3.0\%   &+3.5\%   &+22\%    &+23\% \\
\quad rms          & 2.6\%   & 2.3\%   &4.0\%    &4.1\% \\    
Ge \\
\quad min          &$-$2.2\%   &$-$2.1\%   &$-$11\%    &$-$15\% \\
\quad max          &+3.1\%   &+3.5\%   &+22\%    &+22\% \\
\quad rms          & 2.8\%   & 2.5\%   &4.0\%    &4.1\% \\
\end{tabular}
\end{table}

In order to gain some further insight on the difference between the
SP and DP structures, we look at the maximum deviations of bond
lengths and bond angles for the LDA geometries, as shown in Table II.
The main trend as a function of the material considered is the
smaller variation of bond angles, and greater variation of bond
lengths, in C relative to Si and Ge.  This is consistent with the
fact that the ratio of bond-angle bending to bond-stretching forces
is bigger in C, compared to Si and Ge.
However, we are mainly interested in the trends in going from the
SP to the DP structure.  For all three materials, we find that the
pattern of deviations looks surprisingly similar for the two structures.
In particular, the maximum and minimum bond-length and bond-angle
variations are {\it not} systematically smaller in the DP structure,
in spite of its lower energy.  However, a trend becomes more visible
when inspecting the root-mean-square deviations: we see that the
rms bond-length variations are systematically slightly larger in
the SP core.  While it also appears that the bond-angle variations
go the other way, being slightly smaller for the SP core, this seems
to be a weaker effect.
These results suggest that advantage of the DP structure is that it
allows a better packing of the atoms, as measured by bond-length
variations, at the expense of a slightly greater bond-bending strain.

In view of the fact that the variations in bond lengths and bond
angles between the SP and DP structures are so subtle, it is not at
all clear whether a Keating model could be expected to reproduce
the correct trends in $E_{\rm SP}$ vs.~$E_{\rm DP}$ for the three
materials.  In order to test this, we also present in Table I the
energies for the two core structures as computed using the Keating
potential.  This was done both by consistently relaxing and
evaluating the energy using the Keating model, and by simply
evaluating the Keating energy of the LDA-relaxed supercells.  Note
that in the case of C, the DP-SP energy difference for the
Keating-relaxed geometry is substantially smaller than the
corresponding LDA result. Also, this set of numbers predicts an
incorrect trend for this quantity among the three materials, as
compared to the LDA results. For the larger supercell this quantity
is even in qualitative disagreement with the TETB result, in the
case of C. (We evaluated the Keating energy for the TETB geometries
in these larger cells, and found the correct qualitative behavior,
with $E_{\rm DP}-{\overline E}_{\rm SP} = -85$ meV/\AA\ for C.)  On
the other hand, for the LDA geometries, the Keating results follow
the same trends as those obtained with the other methods.
From these results, we conclude that the Keating potential
cannot be trusted to capture the energy difference between the SP
and DP structures on a quantitative (or perhaps even qualitative)
level.

To summarize, we investigated the energetics of the SP and DP core
reconstructions of the 90$^\circ$ partial dislocation in the
homopolar semiconductors, C, Si, and Ge. We find the DP core to be
favored in all three materials, and observe that the energy
difference between the two geometries follows the same trends as
the energy gap and the stiffness in these materials. The nature of
this difference is primarily associated with the local strains
introduced by the the two reconstructions.  The DP structure is
shown to introduce smaller average bond-length deviations, at the
expansion of slightly larger average bond-angle bending
distortions, with respect to the SP one, with the delicate balance
between these two strain components favoring the DP
reconstruction.

\acknowledgments 
Partial support was provided by the DoD Software
Initiative.  J.B. and D.V. acknowledge support from NSF Grant
DMR-96-13648.

\end{document}